\documentclass[12pt]{article}
\usepackage{epsfig, epsf, graphicx, subfigure}
\usepackage{pstricks, pst-node, psfrag}
\usepackage{amssymb,amsmath,bm}
\usepackage{verbatim,enumerate}
\usepackage{rotating, lscape}
\usepackage{setspace}
\usepackage{enumitem}
\usepackage{booktabs}
\usepackage{mdframed}
\usepackage[flushleft]{threeparttable}

\usepackage{epstopdf}
\usepackage[T1]{fontenc}
\usepackage{tabularx,ragged2e,booktabs,caption}

\usepackage[english]{babel}
\usepackage[round,sort&compress]{natbib}
\usepackage{multirow}
\usepackage{theorem}
\usepackage{float}
\def\00{\mathrm{0}}

\begin{document}

\thispagestyle{empty} \baselineskip=28pt \vskip 5mm
\begin{center} {\LARGE{\bf Multivariate Functional Data Visualization and Outlier Detection}}
\end{center}

\baselineskip=12pt \vskip 5mm

\begin{center}\large
Wenlin Dai and Marc G. Genton\footnote[1]{
\baselineskip=10pt Statistics Program,
King Abdullah University of Science and Technology,
Thuwal 23955-6900, Saudi Arabia.
E-mail: wenlin.dai@kaust.edu.sa, marc.genton@kaust.edu.sa\\
This research was supported by the
King Abdullah University of Science and Technology (KAUST).}
\end{center}

\baselineskip=17pt 
\vskip 5mm \centerline{\today} \vskip 5mm

\begin{center}
{\large{\bf Abstract}}
\end{center}

This paper proposes a new graphical tool, the magnitude-shape (MS) plot, for visualizing both the magnitude and shape outlyingness of multivariate functional data.
The proposed tool builds on the recent notion of functional directional outlyingness, which measures the centrality of functional data by simultaneously considering the level and the direction of their deviation from the central region.
The MS-plot intuitively presents not only levels but also directions of magnitude outlyingness on the horizontal axis or plane, and demonstrates shape outlyingness on the vertical axis.
A dividing curve or surface is provided to separate non-outlying data from the outliers.
Both the simulated data and the practical examples confirm that the MS-plot is superior to existing tools for visualizing centrality and detecting outliers for functional data.

\vskip 12pt

\par\medskip\noindent
{\bf Short title:} Multivariate Functional Data Visualization

\baselineskip=26pt

\vskip 12pt
\section{Introduction}

Thanks to the rapid evolution of technology, data are frequently obtained as trajectories or images in many scientific areas, including but not limited to meteorology, biology, medicine, and engineering.
As a result, functional data analysis has attracted an increasing number of researchers during the past two decades.
Various methods for modeling, clustering, or drawing inferences about functional data have been proposed \citep{ramsayfunctional,ferraty2006nonparametric,horvath2012inference}.

Data visualization is an effective way to explicitly illustrate the characteristics that are not apparent from the mathematical models or summary statistics.
For point-type data, graphical tools such as histograms \citep{pearson1895contributions} and boxplots \citep{tukey1975mathematics} are widely used to intuitively and informatively demonstrate important features of a data set.
For functional data, graphical tools are mainly proposed for univariate cases, including functional bagplots and functional highest density region (HDR) plots \citep{hyndman2010rainbow}, functional boxplots \citep{sun2011functional}, surface boxplots \citep{genton2014surface}, outliergrams \citep{arribas2014shape}, and amplitude and phase boxplot displays \citep{xie2017geometric}.
But multivariate functional data are also quite often observed, e.g., daily wind data at different weather stations, measures of several economic indexes over time \citep{Chowdhury2016}, and images from a video file.
\citet{hubert2015multivariate} and \citet{rousseeuw2016measure} proposed the functional outlier map (FOM), which was the first attempt to provide a graphical tool for multivariate functional data.
In this paper, we aim to contribute to the functional data analysis toolbox by proposing a graphical method, named \emph{magnitude-shape (MS) plot}, which applies to multivariate functional data that have one-dimensional or multi-dimensional support spaces. In designing such a tool, we have followed the principles suggested by \citet{robbins2012creating} and \citet{wickham2010layered}. Comprehensive graphical examples for different types of data are well studied in \citet{unwin2015graphical}.
Also, thanks to the comprehensive guidelines on designing interactive graphics \citep{cook2007interactive} and the development of graphical tools such as, {\em plotly} \citep{plotly} and {\em shiny} \citep{shiny}, we constructed some interactive examples of our MS-plot. Using the R package {\em ggplot2} \citep{ggplot}, we manage to illustrate examples and numerical results much better.

The MS-plot is built on the framework of functional directional outlyingness proposed by \citet{dai2018directional} that, for the first time, measures the centrality of functional data by simultaneously considering both the level and the direction of their deviation from the central region.
Our proposed tool concentrates on visualizing functional data with regard to both magnitude and shape centrality.
In the case of contaminated data sets, an adequate criterion is used to flag various types of outliers that could lead to severe biases in the modeling and forecasting of functional data.

The remainder of this paper is organized as follows. In Section 2, we briefly review the framework of functional directional outlyingness. In Section~3, we propose the MS-plot and some related tools, e.g., the MS-plot array and the multivariate outliergram.
In Section~4, we numerically compare the MS-plot with some existing graphical tools in terms of its capability to indicate the centrality and detect outliers of functional data.
We demonstrate the performance of the proposed tools with two practical applications in Section 5.
We end the paper with a discussion in Section 6.

\vskip 24pt
\section{Functional Directional Outlyingness}
Directional outlyingness \citep{dai2018directional} is a framework that adds direction to the conventional concept of outlyingness, recognizing that the direction of outlyingness is crucial to describing the centrality of multivariate functional data.

Specifically, directional outlyingness for point-type data, is defined as
$$\mathbf{O}(\mathbf{Y},F_{\mathbf{Y}})=\left\{{1/ d(\mathbf{Y},F_{\mathbf{Y}})}-1\right\}\cdot \mathbf{v}, \quad d(\mathbf{Y},F_{\mathbf{Y}})>0,$$
where $F_{\mathbf{Y}}$ denotes the distribution of a random variable $\mathbf{Y}$, $d$ is a conventional depth notion, and $\mathbf{v}$ is the unit vector pointing from the median of $F_{\mathbf{Y}}$ to $\mathbf{Y}$. Assuming that $\mathbf{Z}$ is the unique median of $F_{\mathbf{Y}}$ for the depth notion $d$, $\mathbf{v}$ can be expressed as $\mathbf{v}=\left(\mathbf{Y}-\mathbf{Z}\right)/\|\mathbf{Y}-\mathbf{Z}\|$, where $\|\cdot\|$ denotes the $L_2$ norm.

Suppose $\mathbf{X}$, a $p$-dimensional function defined on a domain $\mathcal{I}$, is from a distribution of functional data, $F_{\mathbf{X}}$.
At each fixed design point $t$ in $\mathcal{I}$, we denote the distribution of $\mathbf{X}(t)$ by $F_{\mathbf{X}(t)}$ with a dimension $p$.
Then the three measures of directional outlyingness for functional data \citep{dai2018directional} are defined as:\\
1) Mean directional outlyingness {\rm (\textbf{MO})},
$$\mathbf{MO}(\mathbf{X},F_{\mathbf{X}})=\int_{\mathcal{I}} \mathbf{O}(\mathbf{X}(t),F_{\mathbf{X}(t)}) w(t){\rm d}t;$$
2) Variation of directional outlyingness {\rm (VO)},
$${\rm VO}(\mathbf{X},F_{\mathbf{X}})=\int_{\mathcal{I}} \|\mathbf{O}(\mathbf{X}(t),F_{\mathbf{X}(t)})-{\rm \mathbf{MO}}(\mathbf{X},F_{\mathbf{X}})\|^2w(t){\rm d}t;$$
3) Functional directional outlyingness {\rm ({FO})},
$${\rm FO}(\mathbf{X},F_{\mathbf{X}})=\int_{\mathcal{I}} \|\mathbf{O}(\mathbf{X}(t),F_{\mathbf{X}(t)})\|^2w(t){\rm d}t,$$
where $w(t)$ is a weight function defined on $\mathcal{I}$, which can be constant or proportional to the local variation at each design point \citep{claeskens2014multivariate}.
Throughout this paper, we use a constant weight function, $w(t)=\{\lambda(\mathcal{I})\}^{-1}$, where $\lambda(\cdot)$ represents the Lebesgue measure. The three measures of outlyingness can be linked with the following simple equation,
\begin{eqnarray}
{\rm FO}=\|\mathbf{MO}\|^2+{\rm VO}. \label{decomposition}
\end{eqnarray}
Equation (\ref{decomposition}) decomposes the total functional outlyingness (FO) into two terms:  the quantity of magnitude outlyingness ($\|\mathbf{MO}\|$) and the quantity of shape outlyingness (VO).
This decomposition provides great flexibility for describing the centrality of functional data and for diagnosing potentially abnormal curves.

Different types of depth and outlyingness notions can be utilized to construct the functional directional outlyingness.
We can divide these depths according to their dependence on either rank or distance information.
The rank-dependent depths include, among others, half-region depth \citep{tukey1975mathematics} and simplicial depth \citep{liu1990notion}; the distance-dependent depths include Mahalanobis depth \citep{mahalanobis1936generalized}, spatial depth \citep{vardi2000multivariate}, and projection depth \citep{zuo2003projection}, among others.
\citet{zuo2000general} described more point-type depths and their detailed classification.
Throughout this paper, we mainly use the projection depth to construct functional directional outlyingness. Specifically, the projection depth is defined as:
${\rm PD}(\mathbf{X}(t),F_{\mathbf{X}(t)})=\{1+{\rm SDO}(\mathbf{X}(t),F_{\mathbf{X}(t)})\}^{-1}$, where
$${\rm SDO}(\mathbf{X}(t),F_{\mathbf{X}(t)})=\sup_{\|\mathbf{u}\|=1}{\|\mathbf{u}^{\rm T}\mathbf{X}(t)-{\rm median}(\mathbf{u}^{\rm T}\mathbf{X}(t))\|/ {\rm MAD}(\mathbf{u}^{\rm T}\mathbf{X}(t))}$$
is the Stahel-Donoho outlyingness \citep{stahel1981breakdown,don0ho_1982breakdown} and MAD denotes the median absolute deviation.

\vskip 12pt
\section{Graphical Tools}
This section provides three graphical tools: the MS-plot, the MS-plot array, and the bivariate outliergram induced by directional outlyingness.

\vskip 5pt
\subsection{The magnitude-shape plot}

The MS-plot is a scatter plot of points, $(\mathbf{MO}^{\rm T},{\rm VO})^{\rm T}$, for a group of functional data.
This tool can be used to illustrate the centrality of curves with a response space up to two dimensions.
From the definitions of $\mathbf{MO}$ and ${\rm VO}$, we may expect that the central curves are mapped to the lower central region (small $\|\mathbf{MO}\|$ and small ${\rm VO}$) of the MS-plot.
Shifted outliers are mapped to the lower-outer region (large $\|\mathbf{MO}\|$ and small ${\rm VO}$) of the MS-plot, and the different directions of $\mathbf{MO}$ indicate the different directions of their shifts.
Isolated outliers, that are outlying in a small part of the support space, are mapped to the upper central region (small $\|\mathbf{MO}\|$ and large ${\rm VO}$).
Points in the upper-outer region (large $\|\mathbf{MO}\|$ and large ${\rm VO}$) correspond to the curves that are substantially outlying in both magnitude and shape.

\begin{figure}[t!]
\begin{center}
\includegraphics[width=15cm,height=15cm]{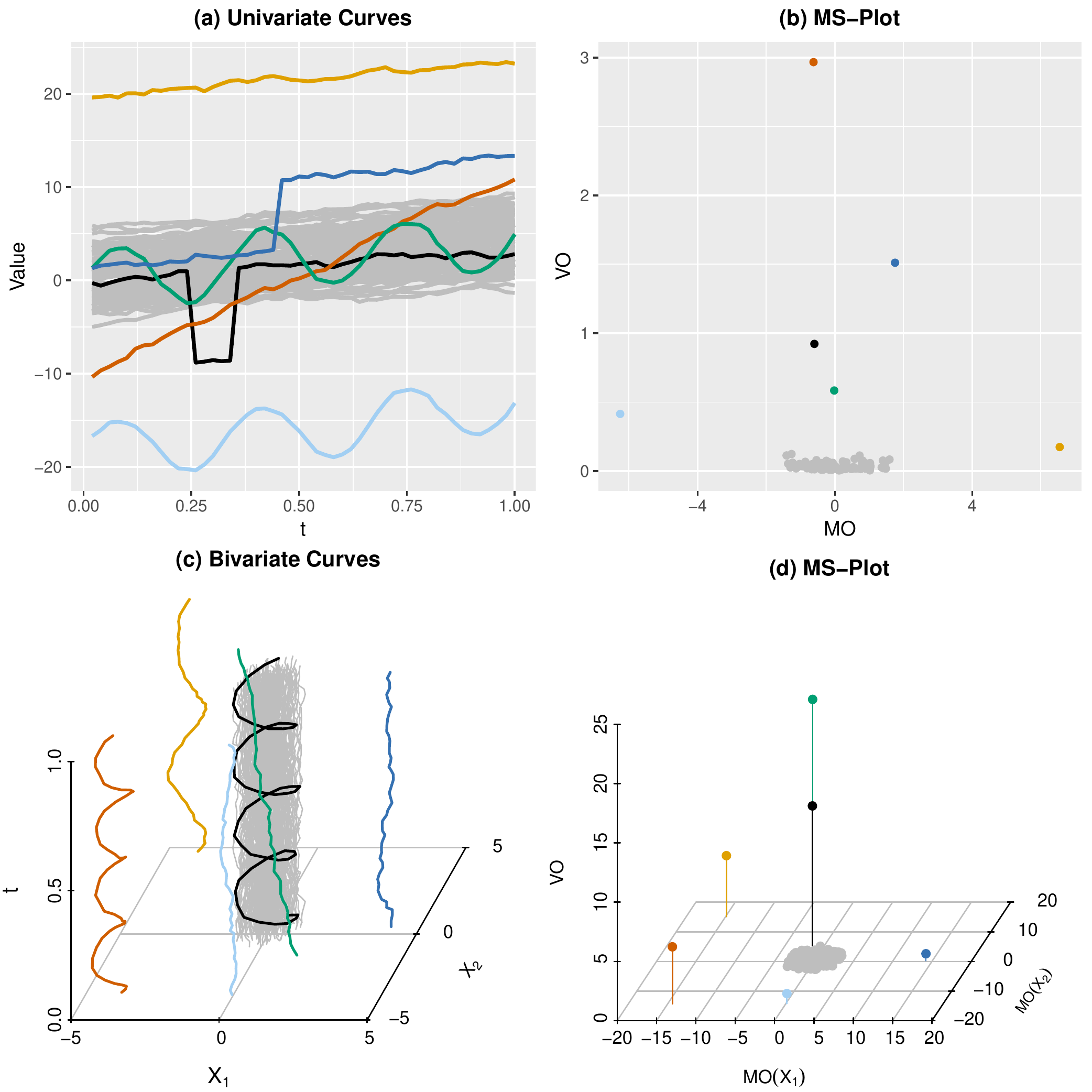}
\caption{Upper panels: a group of univariate curves with various types of outliers (left) and its MS-plot (right). Lower panels: a group of bivariate curves with various types of outliers (left) and its MS-plot (right). {Curves are one-to-one projected as points that reveal the centrality of their corresponding curves.}}
\label{Example_simu}
\end{center}
\end{figure}

To illustrate this, we generate the MS-plot for two simulated data sets, one for univariate and the other for bivariate curves.
Consider a univariate example that consists of 94 non-outlying (grey) curves and six different types of outliers (color) in Figure \ref{Example_simu}(a).
We present the MS-plot for these curves in Figure \ref{Example_simu}(b).
Various types of functional outliers commonly observed are investigated, including shifted outliers, isolated outliers, shape outliers, and some combination of these.
First of all, all six outliers are not only clearly isolated from the grey points, but also differ from each other significantly.
Specifically, the orange point possesses a large ${\rm MO}$ but a small ${\rm VO}$, which is consistent with the fact that the orange line is a shifted outlier;
in the upper middle region of this plot are the four points possessing moderate {\rm MO} but large {\rm VO}, which corresponds to their shape outlyingness;
the cyan point in the middle left of the plot possesses a small ${\rm MO}$ and a relatively large ${\rm VO}$, which coincides with its outlyingness in both magnitude and shape.
Our findings are similar for the bivariate example in the second row of Figure \ref{Example_simu}.
The above two examples illustrate that the MS-plot is capable of visualizing the centrality of functional data and, consequently, can handle various types of outliers effectively.

When the dimension is higher than two, we provide two alternatives for the MS-plot. One option is to plot $(\|\mathbf{MO}\|,{\rm VO})^{\rm T}$ instead of $(\mathbf{MO}^{\rm T},{\rm VO})^{\rm T}$, which presents the overall magnitude outlyingness and shape outlyingness without direction information. 
{To retain the direction information, another option is the parallel coordinate plots \citep{cook2007interactive}. We provide two illustrative plots in the supplemental material.}

\vskip 5pt
\subsection{The magnitude-shape plot array}

\begin{figure}[b!]
	\begin{center}
		\includegraphics[width=15cm,height=15cm]{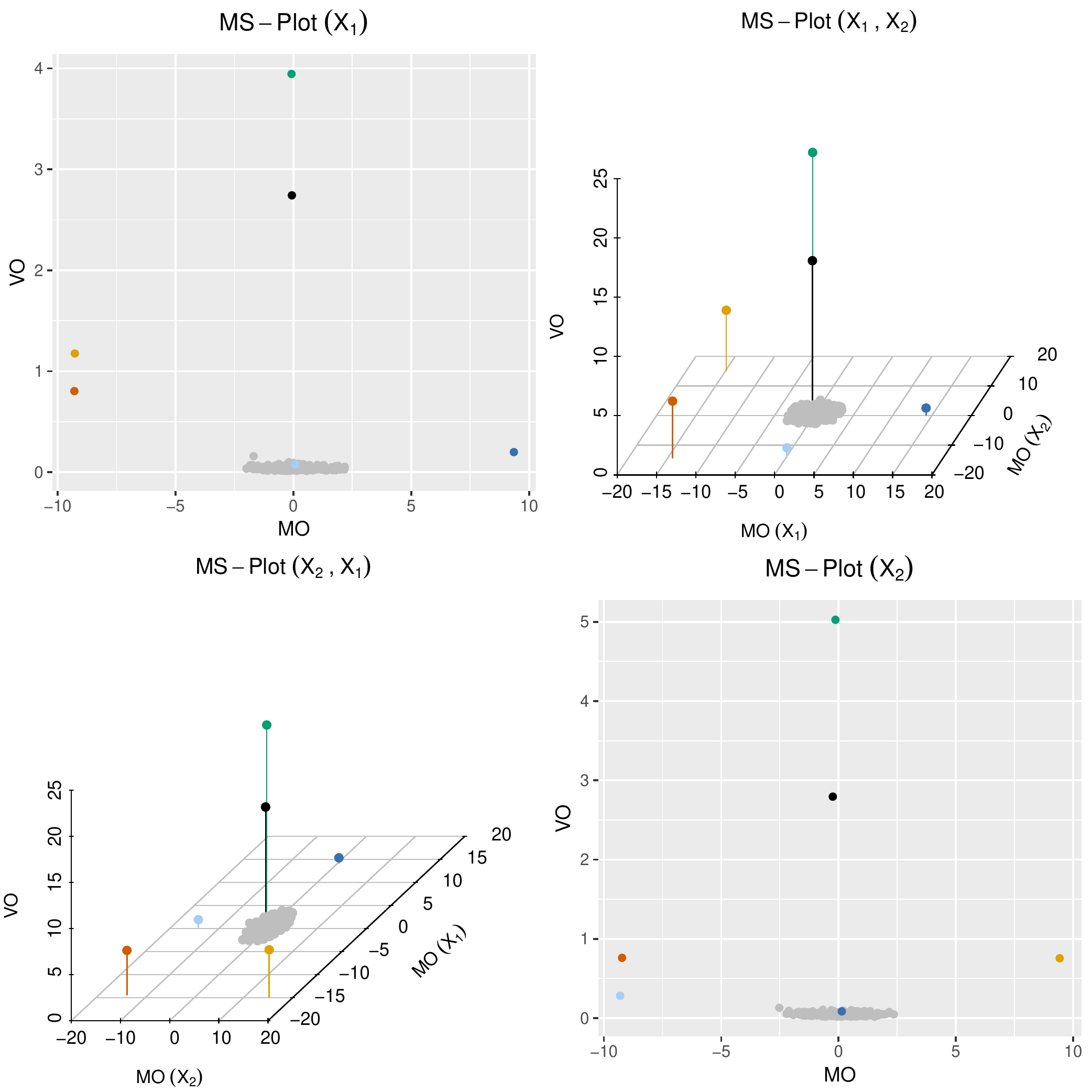}
		\caption{MS-plot array for the bivariate curves from Figure \ref{Example_simu}. Diagonal: the marginal MS-plots; off-diagonal: the pairwise joint MS-plots. {The MS-plot array is an alternative to the MS-plot when the dimension of functional data is two or higher.}}
		\label{MS-Matrix}
	\end{center}
\end{figure}

We provide the MS-plot array to comprehensively illustrate the centrality of multivariate functional data in a single figure.
An example of the MS-plot array is generated for the bivariate curves in Figure \ref{Example_simu} and demonstrated in Figure \ref{MS-Matrix}.

As shown, the MS-plot of the $k$th component is presented on the $k$th diagonal position of the array, and the MS-plot of two combined components (bivariate curves) is presented in the corresponding off-diagonal position.
Among the raw curves exist six outliers, shown in the two joint plots (off-diagonal); however, only five outliers are presented in the two marginal plots (diagonal): the cyan and blue points are missing from these two plots, respectively.
It may happen that an outlier is not outlying for any single component; in this case only a joint analysis of the multivariate curves will reveal its outlyingness.
Hence, treating multivariate functional data jointly instead of marginally is necessary to assess their centrality more comprehensively and accurately.

\begin{figure}[b!]
\begin{center}
\includegraphics[width=16cm,height=8cm]{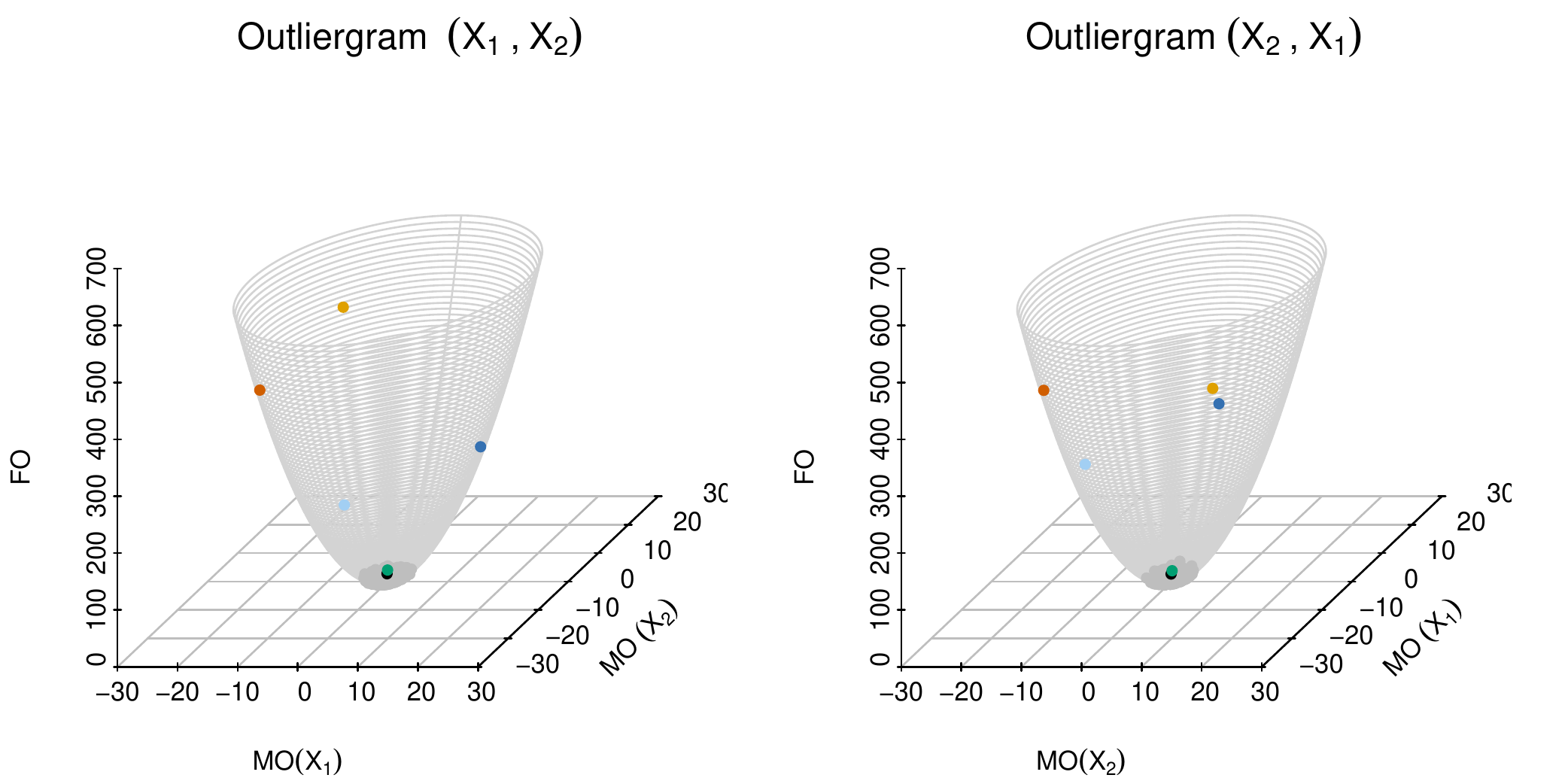}
\caption{Bivariate outliergrams for the bivariate curves from Figure \ref{Example_simu}.}
\label{Multi-outliergram}
\end{center}
\end{figure}

\vskip 5pt
\subsection{The bivariate outliergram}

\citet{arribas2014shape} proposed the outliergram to detect outliers in univariate functional data.
Based on the decomposition (\ref{decomposition}), we now generalize their outliergram to bivariate functional data.
When a group of curves share a common shape, the shape outlyingness ${\rm VO}$ is close to zero by its definition.
Then, a quadratic relationship appears between ${\rm FO}$ and $\|{\rm \mathbf{MO}}\|$.
If we plot the points of $(\mathbf{MO}^{\rm T},{\rm FO})^{\rm T}$, they should be located on a three-dimensional quadratic surface, $\{(a,b,c): a^2+b^2=c,~c>0\}$.
Two examples of bivariate outliergrams are presented in Figure \ref{Multi-outliergram} using the bivariate curves from Figure \ref{Example_simu}.
The magnitude outlyingness is well illustrated, but the shape outlyingness is not as obvious as in the bivariate MS-plot, because it is measured by the vertical distance between the point and the quadratic surface.
For cases with higher dimensions, a two-dimensional plot of $(\|\mathbf{MO}\|,{\rm FO})^{\rm T}$ can be applied, which is the right half of a quadratic curve, because $\|\mathbf{MO}\|\ge 0$.
In such a plot, one can only read the level of the magnitude outlyingness, without the direction. {With the simulation studies in the next section, we show that the outliergram is not as efficient as the MS-plot to visualize the centrality of functional data.
Hence, we recommend using the MS-plot for practical implementations.}

\vskip 12pt
\section{Simulation Studies}

\vskip 5pt
\subsection{Competing methods}

Here, we review three of the graphical tools that have been proposed for illustrating the centrality of functional data: functional principal component analysis (FPCA) \citep{hyndman2010rainbow}, the outliergram \citep{arribas2014shape}, and the functional outlier map (FOM) \citep{rousseeuw2016measure}.
In this section, we compare our MS-plot with the three tools above, in terms of their capability to visualize centrality and detect outliers in functional data analysis.

\textbf{FPCA} Robust principal component analysis based on projection pursuit \citep{croux2005high} is applied to decompose the discretized curves. 
A plot of the first two principal component scores is used to illustrate the centrality of functional data by \cite{hyndman2010rainbow}, who also proposed two graphical tools, the functional bagplot and the functional HDR plot, to identify outliers. {To caculate the scores, we utilize the functions in the R package {\em pcaPP}} \citep{pcaPP}.

\textbf{Outliergram} A quadratic connection is built between the modified band depth (MBD) \citep{lopez2009concept} and the modified epigraph index (MEI) \citep{lopez2011half}. A plot of the MBD against the MEI is used to show the centrality of univariate curves. A combination of the adjusted functional boxplot \citep{sun2012adjusted} and the outliergram is used to detect magnitude and shape outliers simultaneously.
As discussed in \citet{dai2018directional}, the connection between the MBD and the MEI fits exactly into the decomposition (\ref{decomposition}) induced by the framework of functional directional outlyingness.
{To caculate MBD and MEI, we utilize the functions in the R package {\em roahd}} \citep{roahd}.

\textbf{FOM} Skewness-adjusted outlyingness is defined for point-type data first; then, the functional skewness-adjusted outlyingness (fAO) and the variability of point-wise skewness-adjusted outlyingness (vAO) are defined for functional data \citep{rousseeuw2016measure}.
A plot of ${\rm (fAO,vAO)}^{\rm T}$ is proposed to visualize the centrality of the functional data.
{To caculate fAO and vAO, we utilize the functions provided by the authors on the website  \textcolor[rgb]{0,0,1.00}{https://wis.kuleuven.be/stat/robust/software}, which now have been included in the R package {\em mrfDepth}} \citep{mrfDepth}.
Actually, \citet{rousseeuw2016measure} also used the term ``directional outlyingness''.
Unlike our directional outlyingness, their definition does not account for the direction, in which a curve deviates from the central region, and is always a scalar no matter if the curves are univariate or multivariate.
To avoid confusion, we call their proposed quantities ``skewness-adjusted outlyingness'', which is the essential idea of their definition.
Because the direction of outlyingness is not accounted for, the FOM is not as informative as the MS-plot.

\vskip 5pt
\subsection{Simulation design}

To compare our MS-plot with the three existing tools above and assess their performance for both centrality visualization and outlier detection, we consider the following four models for introducing either magnitude outliers or shape outliers:

\begin{itemize}[noitemsep]
\item [ ] \textbf{Model 1 (shifted outlier)} \\
Main model: $X(t)=4t+e(t)$, \\
Contamination model: $X(t)=4t+8U+e(t)$, \\
for $0\le t\le1$, where $e(t)$ is a Gaussian process with zero mean and covariance function $\gamma(s,t)=\exp\{-|t-s|\}$, and $U$ takes values $-1$ and $1$ with probability $1/2$. The contaminating curves shift up and down from the main model.
\item [ ] \textbf{Model 2 (isolated outlier)} \\
Main model: $X(t)=4t+e(t)$, \\
Contamination model: $X(t)=4t+8UI_{\{T\le t\le T+0.05\}}+e(t)$, \\
for $0\le t\le1$, where $T$ is generated from a uniform distribution on $[0.1,0.9]$ and $I_{A}$ is an indicator function taking value 1 on the set $A$ and 0 otherwise. The contaminating curves add spikes to the main model.
\item [ ] \textbf{Model 3 (shape outlier I)} \\
Main model: $X(t)=30t(1-t)^{3/2}+{\tilde e}(t)$,\\
Contamination model: $X(t)=30(1-t)t^{3/2}+\tilde{e}(t)$,\\
for $0\le t\le1$, where ${\tilde{e}}(t)$ is a Gaussian process with zero mean and covariance function $\tilde\gamma(s,t)=0.3\exp\{-|t-s|/0.3\}$. The contaminating curves are defined on the reversed time interval of the main model.
\item [ ] \textbf{Model 4 (shape outlier II)} \\
Main model: $X(t)=4t+\tilde e_1(t)$, \\
Contamination model: $X(t)=4t+\tilde e_2(t)$, \\
for $0\le t\le1$, where $\tilde e_1(t)$ is a Gaussian process with zero mean and covariance function $\gamma_1(s,t)=\exp\{-|t-s|\}$, and $\tilde e_2(t)$ is a Gaussian process with zero mean and covariance function $\gamma_2(s,t)=5\exp\{-2|t-s|^{0.5}\}$. The contaminating curves share the same trend as the main model, but have different covariance functions.
\end{itemize}

\begin{figure}[t!]
	\begin{center}
		\includegraphics[width=16cm,height=4cm]{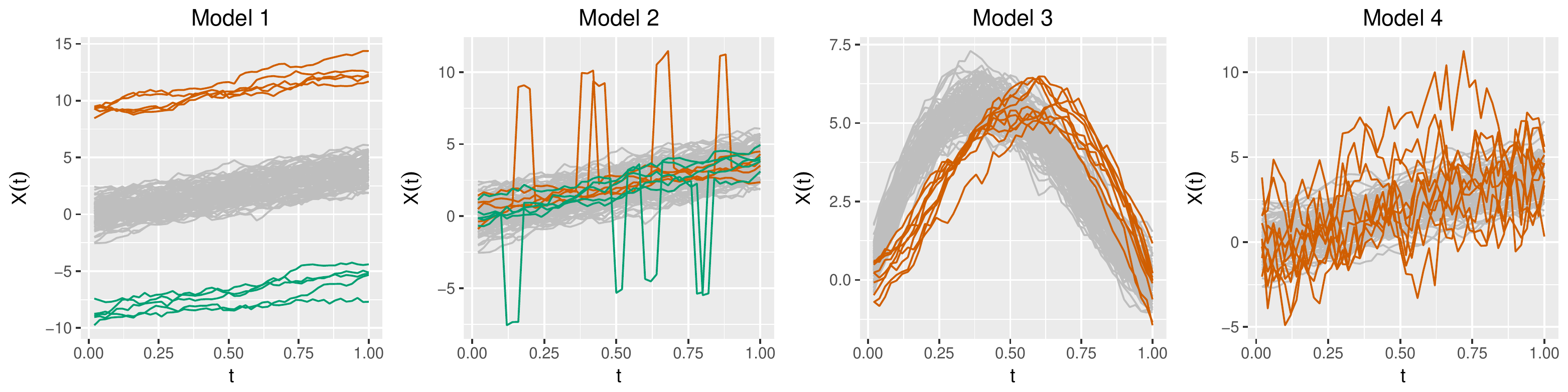}
		\caption{Illustrative examples of the four univariate models. Each plot displays a sample of size 100, with 90 curves (grey) from the main model and 10 outliers (red or green) from the contamination models.}
		\label{simu_settings}
	\end{center}
\end{figure}

We generate samples from the four models with sample size $n=100$ and contamination level $c=0.1$, and illustrate them in Figure \ref{simu_settings}.

\vskip 5pt
\subsection{Centrality visualization}

\begin{figure}[t!]
	\begin{center}
		\includegraphics[width=16cm,height=16cm]{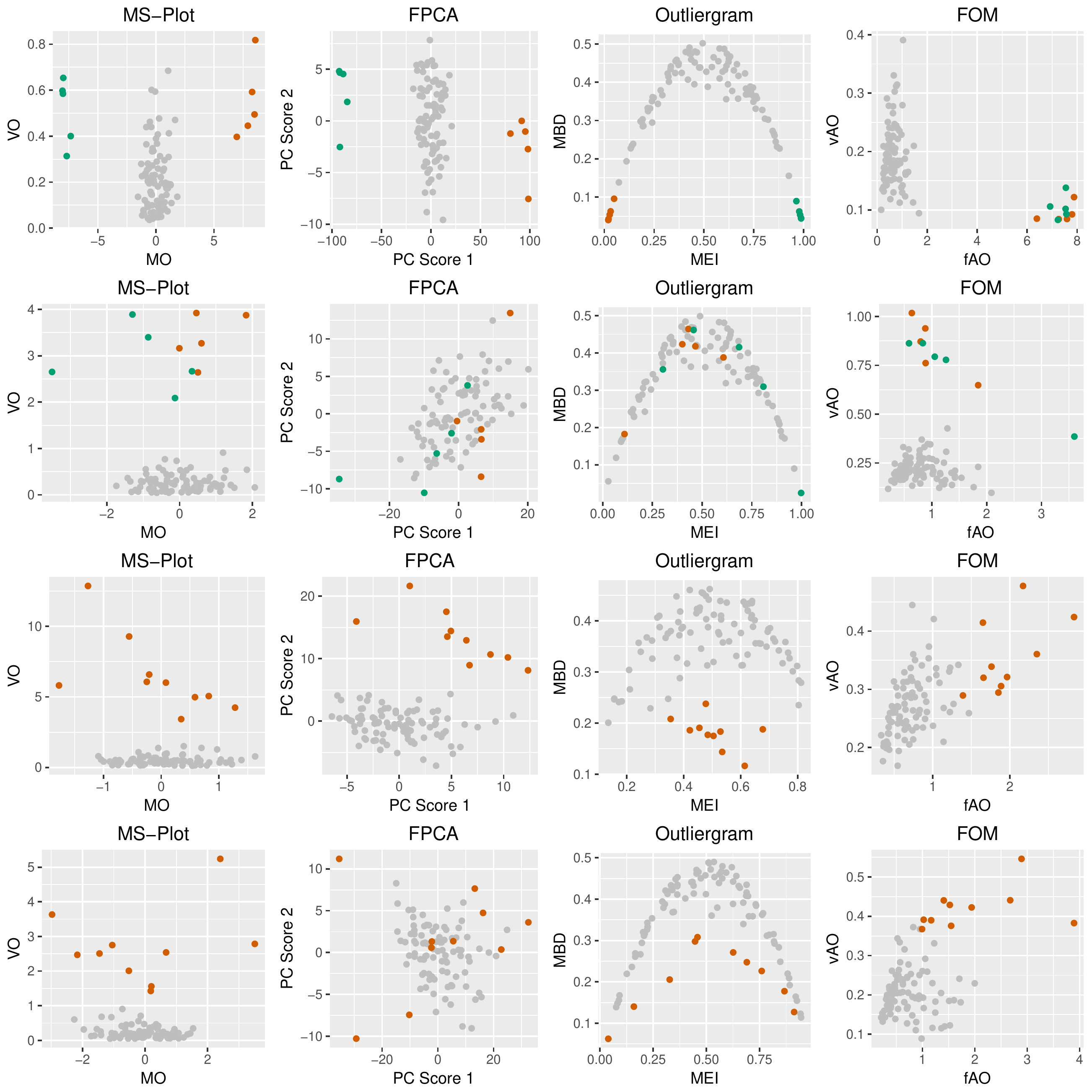}
		\caption{Visualization results of the four tools for the centrality of four groups of univariate curves. First Row: Model 1; second row: Model 2; third row: Model 3; fourth row: Model 4. First column: the MS-plot; second column: FPCA; third column: Outliergram; fourth column: the FOM. Colors follow the settings in Figure \ref{simu_settings}. {The MS-plot outperforms the other three tools by visualizing the centrality of curves more efficiently.}}
		\label{central_visual}
	\end{center}
\end{figure}
We first compare the four graphical tools through visualizing the centrality of functional data.
An excellent graphical tool should be able to show the centrality of both magnitude and shape, which are two important features of a curve.
Also, we expect the tool to be applicable to functional data with broader types of structures, i.e., both the support space and the response space can be either one-dimensional or multi-dimensional.
We apply the four methods to the contaminated samples from Figure~\ref{simu_settings} and illustrate the results in Figure~\ref{central_visual}.

For Model 1, which is contaminated by ten shifted outliers, the four methods handle the magnitude outlyingness well, and all but the FOM further demonstrate the direction of magnitude outlyingness.
Specifically, both the level and the direction of magnitude outlyingness are indicated by the MO in the MS-plot, by the first PC score in FPCA, and by the MEI in the outliergram, respectively; however, the ${\rm fAO}$ in the FOM only captures the levels of magnitude outlyingness.
The red and green points are utilized to distinguish the upward and downward shifted outliers; they are located in different directions on the MS-plot, FPCA, and the outliergram, but in the same direction on the FOM.
A similar discrepancy appears in the second row of Figure \ref{central_visual} for Model 2, which is contaminated by 10 isolated outliers.

For Models 2--4 containing shape outliers, both the MS-plot and the FOM illustrate the shape outlyingness well through VO and vAO, respectively.
However, FPCA and the outliergram fail to show the shape outlyingness, especially when the shape outliers do not deviate from the non-outlying curves in the same fashion, e.g., Models 2 and 4.
Moreover, the MS-plot separates the shape outliers from the normal curves much better than the FOM for Model 3 because it accounts for the direction of outlyingness.

Finally, the conventional outliergram applies only to univariate functional data.
The functional PCA is widely applied to analyze multivariate functional data, but a similar two-dimensional plot of the PC scores is not available to effectively describe the centrality of the underlying multivariate curves or images.
In contrast, both the MS-plot and the FOM apply to multivariate functional data defined in either one-dimensional or multi-dimensional spaces.
For clarity, we summarize the above comparison results in Table \ref{comparison_result}.

\begin{table}[H]
	\caption{Comparison of the four graphical tools: MS-plot, FPCA, outliergram, and FOM.}
	\begin{center}
		\begin{tabular}{rcccccccc}   \toprule[1pt]
			&&    MS-Plot&& FPCA && Outliergram && FOM \\ \hline
			Direction of magnitude outlyingness           &&    $\surd$    &&  $\surd$    && $\surd$   && $\times$ \\ \hline
			Shape outlyingness           &&    $\surd$    &&  $\times$   && $\times$  && $\surd$  \\ \hline
			Multivariate functional data           &&    $\surd$    &&  $\times$   && $\times$   && $\surd$  \\
			\bottomrule[1pt]
		\end{tabular}
	\end{center}
	\label{comparison_result}
\end{table}

\vskip 5pt
\subsection{Outlier detection}
The above three tools and our MS-plot all map the functional data to multivariate points.
The centrality of these points corresponds to that of the functional data.
After visualizing the centrality in the first step, the second step is to accurately separate the functional outliers from the non-outlying curves.
All four tools are equipped with their specific outlier detection procedures.
Unlike the other three tools that detect outliers using only their plots, the outliergram is combined with the functional boxplot to find abnormal curves.
FPCA poorly visualizes even the centrality when the outliers are of various types of outlyingness (Models 2 and 4), and so it is unreasonable to expect a satisfying performance in outlier detection from this tool.
Hence, in this subsection, we focus on comparing the MS-plot with the FOM in outlier detection, for both univariate and multivariate functional data.
Note that we do not consider other existing outlier detection methods that rely on either a model \citep{gervini2009detecting,yu2012outlier} or a cutoff obtained through a bootstrap study of functional depth \citep{febrero2008outlier}.

Under the assumptions that the underlying functional data follow a normal distribution and that the random projection point-wise depth is adopted to calculate functional directional outlyingness, \citet{dai2018directional} proposed an outlier detection method based on $(\mathbf{MO}^{\rm T},{\rm VO})^{\rm T}$.
In particular, they calculated the squared robust Mahalanobis distance (SRMD) for $(\mathbf{MO}^{\rm T},{\rm VO})^{\rm T}$ with the covariance matrix estimated by the minimum covariance determinant (MCD) algorithm \citep{rousseeuw1985multivariate,rousseeuw1999fast}.
Then, they approximated the tail distribution of SRMD with a Fisher's $F$ distribution, according to the procedure suggested by \citet{hardin2005distribution}.
Finally, the curves that caused SRMD to exceed a threshold value obtained from this $F$ distribution were flagged as outliers.
To better demonstrate this outlier detection procedure, we add an ellipsoid, determined by both the MCD covariance matrix and the threshold value, to the MS-plot that separates non-outlying curves from outliers.

For the FOM, \citet{rousseeuw2016measure} defined the following outlier detection criteria. First, they calculated the combined functional outlyingness (CFO) as
$${\rm CFO_i=\sqrt{\{fAO_i/med(fAO)\}^2+\{vAO_i/med(vAO)\}^2}},$$
where ${\rm med(fAO)}$ denotes the median of $\{{\rm fAO}_1,\dots,{\rm fAO}_n\}$ and ${\rm med(vAO)}$ denotes the median of $\{ {\rm vAO}_1,\dots,{\rm vAO}_n\}$. Then, they transformed the ${\rm CFO}$ to ${\rm LCFO=log(0.1+CFO)}$, and eventually flagged a function as an outlier if $\{{\rm LCFO}_i-{\rm med(LCFO)}\}/{\rm MAD(LCFO)}>\Phi^{-1}(0.995)$, where ${\rm \Phi}$ is the standard normal cumulative distribution function.

\begin{figure}[t!]
\begin{center}
\includegraphics[width=16cm,height=16cm]{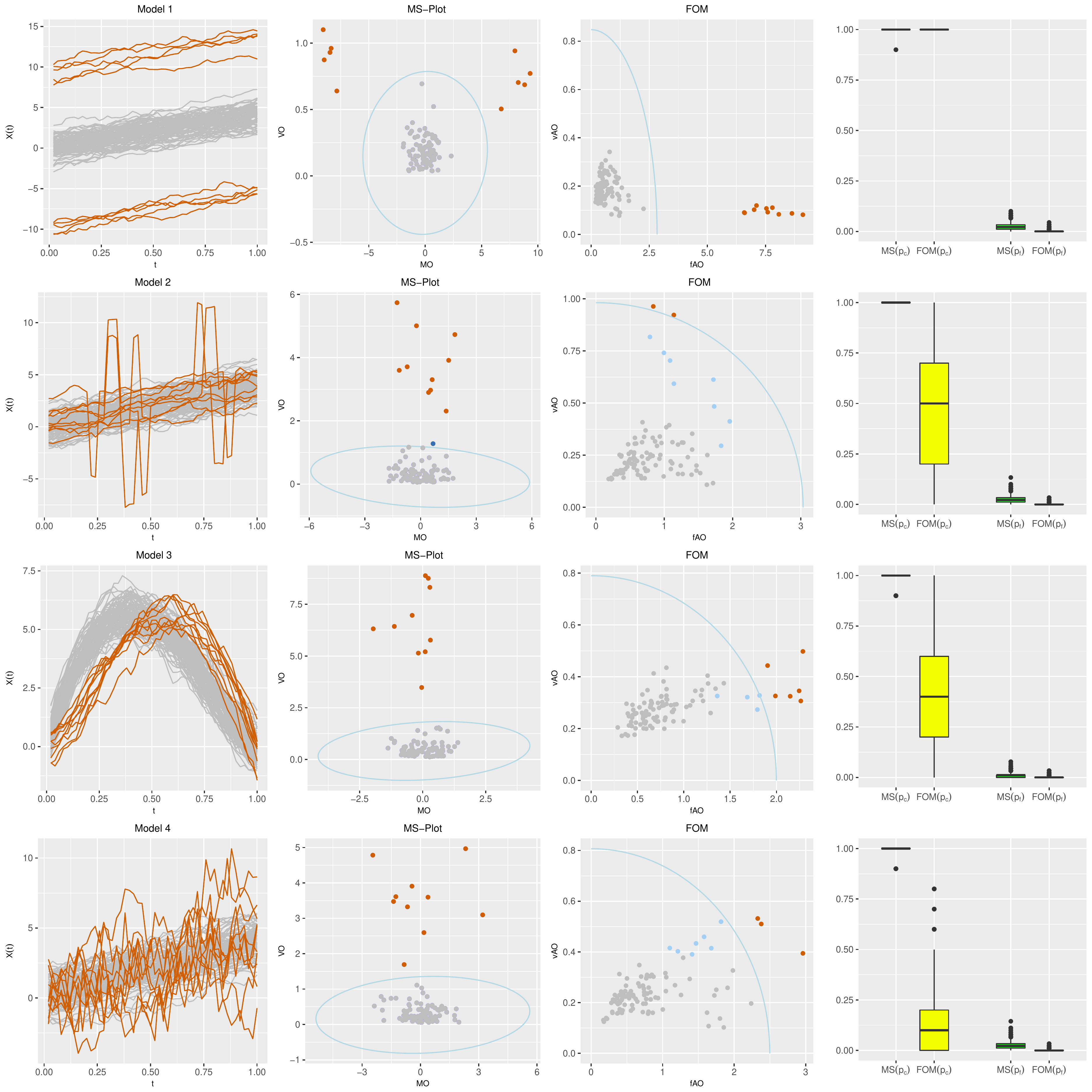}\\
\caption{First column: univariate curves generated from Models 1--4 (from top to bottom). Second and third columns: the outlier detection procedures of the MS-plot and the FOM, where four colors are used to denote different types of points: correctly detected outliers (red); falsely detected outliers (blue); undetected outliers (cyan); true non-outlying curves (grey). Fourth column: numerical results of the MS-plot and the FOM for four contaminated data sets: MS-plot (green) and FOM (yellow). {Although FOM reveals clearly the outlyingness of the abnormal curves, it fails to detect them accurately.}}
\label{outlier_detect}
\end{center}
\end{figure}

Using the four previous examples from Figure 4, we illustrate the outlier detection procedures of the MS-plot and the FOM in the second and third columns of Figure 6, respectively.
For Model 1, both methods identify all the shifted outliers accurately. For Models 2--4, the MS-plot performs much better than the FOM, although they both visualize the outlyingness of outliers quite well.
To numerically evaluate the performance of the two tools, we obtained 1000 replications of the above simulation.
For comparison, we calculated two quantities for each run: the correct detection rate $p_c$ and the false detection rate $p_f$.
We present the numerical results using boxplots in the fourth column of Figure \ref{outlier_detect}.
As shown, both methods detect all the shifted outliers, but the $p_f$ of the MS-plot is a little bit higher.
For the shape outliers, the $p_c$ of the FOM is much smaller than that of the MS-plot, and the $p_f$ of the FOM is lower than that of the MS-plot, which are consistent with the graphical examples in the second and third columns of Figure \ref{outlier_detect}.
Hence, we conclude that the FOM is more sensitive to magnitude outlyingness, i.e., purely shifted outliers; however, the MS-plot is more sensitive to shape outlyingness, which is harder to handle and of more interest to functional data analysis. Also, the outlier detection criterion of the MS-plot is more efficient than that of the FOM.

Both investigated methods also apply to multivariate as well as univariate functional data.
Here, we consider one bivariate setting to show the necessity of considering the direction of outlyingness when describing the centrality of multivariate functional data:
\begin{itemize}[noitemsep]
\item [ ] \textbf{Model 5 (bivariate shape outlier)} \\
Main model: $\mathbf{X}(t)=\mathbf{e}(t)+\mathbf{U}$, \\
Contamination model: $\mathbf{X}(t)=\mathbf{e}(t)+(\sin(4\pi t),\cos(8\pi t))^{\rm T}$, \\
where $\mathbf{U}=(U_1,U_2)^{\rm T}$ and $U_1$, $U_2$ are randomly generated from the uniform distribution on the interval $[-1.1,1.1]$; $\mathbf{e}(t)=\{{\rm e}_1(t), {\rm e}_2(t)\}^{\rm T}$ is a bivariate Gaussian process with zero mean and a cross-covariance function \citep{gneiting2010matern}:
$$C_{ij}(s,t)=\rho_{ij}\sigma_{i}\sigma_{j}\mathcal{M}(|s-t|;\nu_{ij},\alpha_{ij}), \quad i,j=1,2,$$
for $0\le t,s\le 1$, where $\rho_{12}$ is the correlation between $X_1(t)$ and $X_2(t)$, $\rho_{11}=\rho_{22}=1$, $\sigma_i^2$ is the marginal variance, and $\mathcal{M}(h;\nu,\alpha)=2^{1-\nu}\Gamma(\nu)^{-1}\left(\alpha|h|\right)^{\nu}\mathcal{K}_\nu(\alpha|h|)$ with $|h|=|s-t|$, is the Mat{\'e}rn class \citep{matern1960spatial}, where $\mathcal{K}_\nu$ is a modified Bessel function of the second kind, $\nu>0$ is a smoothness parameter, and $\alpha>0$ is a range parameter. Here, we set the following parameter values: $\sigma_1=\sigma_2=0.1$, $\alpha_{11}=0.2$, $\alpha_{22}=0.1$, $\alpha_{12}=0.16$, $\nu_{11}=1.2$, $\nu_{22}=0.6$, $\nu_{12}=1$, and $\rho_{12}=0.1$.
\end{itemize}

\begin{figure}[b!]
	\begin{center}
		\includegraphics[width=16cm,height=12cm]{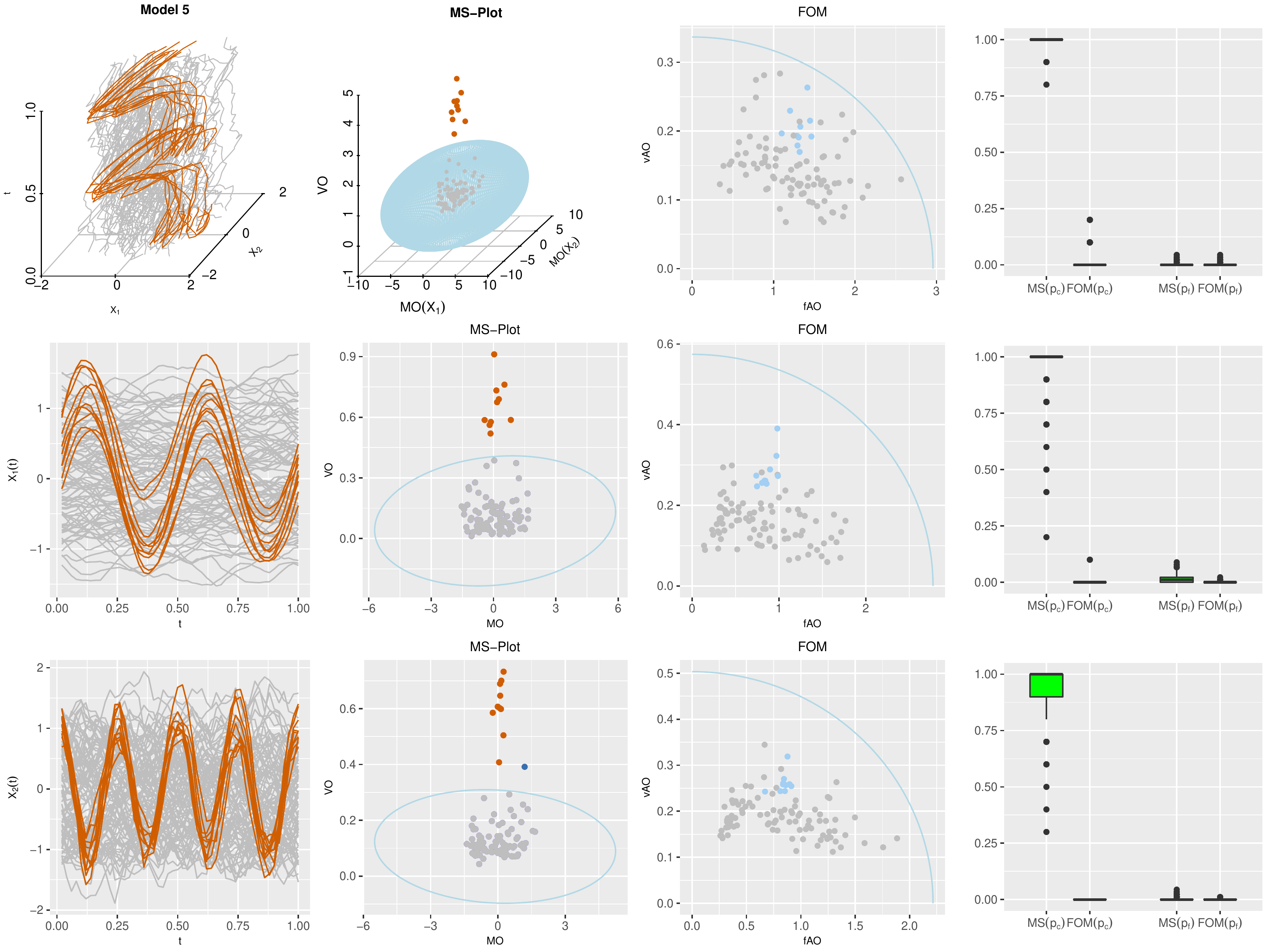}\\
		\caption{First column: bivariate curves generated from Model 5 (top), univariate curves of the first component (middle), and univariate curves of the second component (bottom). Second and third columns: the outlier detection procedures of the MS-plot and the FOM, where four colors are used to denote different types of points: correctly detected outliers (red); falsely detected outliers (blue); undetected outliers (cyan); true non-outlying curves (grey). Fourth column: numerical results of the MS-plot and the FOM for the bivariate curves and the univariate marginal curves: MS-plot (green) and FOM (yellow). {Unlike the MS-plot, FOM fails to visualize the outlyingness of the abnormal curves for these examples.}}
		\label{counter-example}
	\end{center}
\end{figure}

A realization of the above model with sample size $n=100$ and contamination level $c=0.1$ is illustrated in the first column of Figure \ref{counter-example}.
Both the MS-plot and the FOM are used to assess the centrality of the curves in this example, and the results are shown in the second and third columns of Figure \ref{counter-example}.
Unlike the performance for the previous four univariate settings, the FOM ultimately fails to visualize the shape outlyingness of the introduced outliers in Model 5.
The outliers in Model 5 keep a stable level of outlyingness at each time point, so they are outlying only because their directions of outlyingness change across the whole design domain.
However, the FOM (i.e., the skewness-adjusted-outlyingness) does not account for this type of information at all when measuring the centrality of multivariate functional data.
Therefore, in the fourth column of Figure \ref{counter-example}, we are not surprised to see poor performance of the FOM in detecting outliers from Model 5.
For the marginal curves, we obtained similar simulation results, which are again due to the FOM's lack of information about the direction (positive or negative signs for univariate curves) of outlyingness.
Comparing the joint and marginal outlier detection results provided by the MS-plot shows that the joint method performs even better than the two marginal methods for both correct and false detection rates.
Of course, choosing either the marginal or joint MS-plots depends on the purpose of the data analysis. If the users are interested in finding the anomalies in a single component, then the marginal MS-plot of the corresponding component should be utilized; if the users are interested in detecting the anomalies in the interaction of several components, then the joint MS-plot should be employed.

After obtaining ${\rm (\mathbf{MO}^{\rm T}, VO)}^{\rm T}$, besides the squared robust Mahalanobis distance method, other multivariate data outlier detection methods can also be utilized to find the anomalies. 
For instance, we may calculate a second-step depth to measure the centrality of each multivariate point on the plot and use the bootstrap to find a proper cut-off value to detect outliers. Alternatively, one may simply apply the boxplot to each dimension of $\mathbf{MO}$ and VO separately to detect the different types of outliers. Moreover, the cutoff value can be adaptively chosen by changing the inflating factor of the boxplot.

\vskip 12pt
\section{Applications}

Besides simulations, we test our proposed method on two data sets. One data set involves univariate curves and the other one includes bivariate curves.
Applications to functional data with more general structures, such as multivariate curves with more than two dimensions or image data from a video recording, are presented in the Supplemental Material.

\vskip 5pt
\subsection{Tecator data}
The first data set is the Tecator data: 215 near-infrared absorbance spectra of meat samples recorded on a Tecator Infratec Food Analyzer with a wavelength range of 850--1050 nm, which is loaded from the R package \emph{fda.usc}.
Each observation consists of a 100-channel absorbance spectrum within this range.
The percentages of water, fat, and protein content are also available for each sample.
We first divide the samples into two parts according to their fat content: higher than $20\%$ (77 samples, blue) and lower than $20\%$ (138 samples,~red).

\begin{figure}[b!]
\begin{center}
\includegraphics[width=16cm,height=8cm]{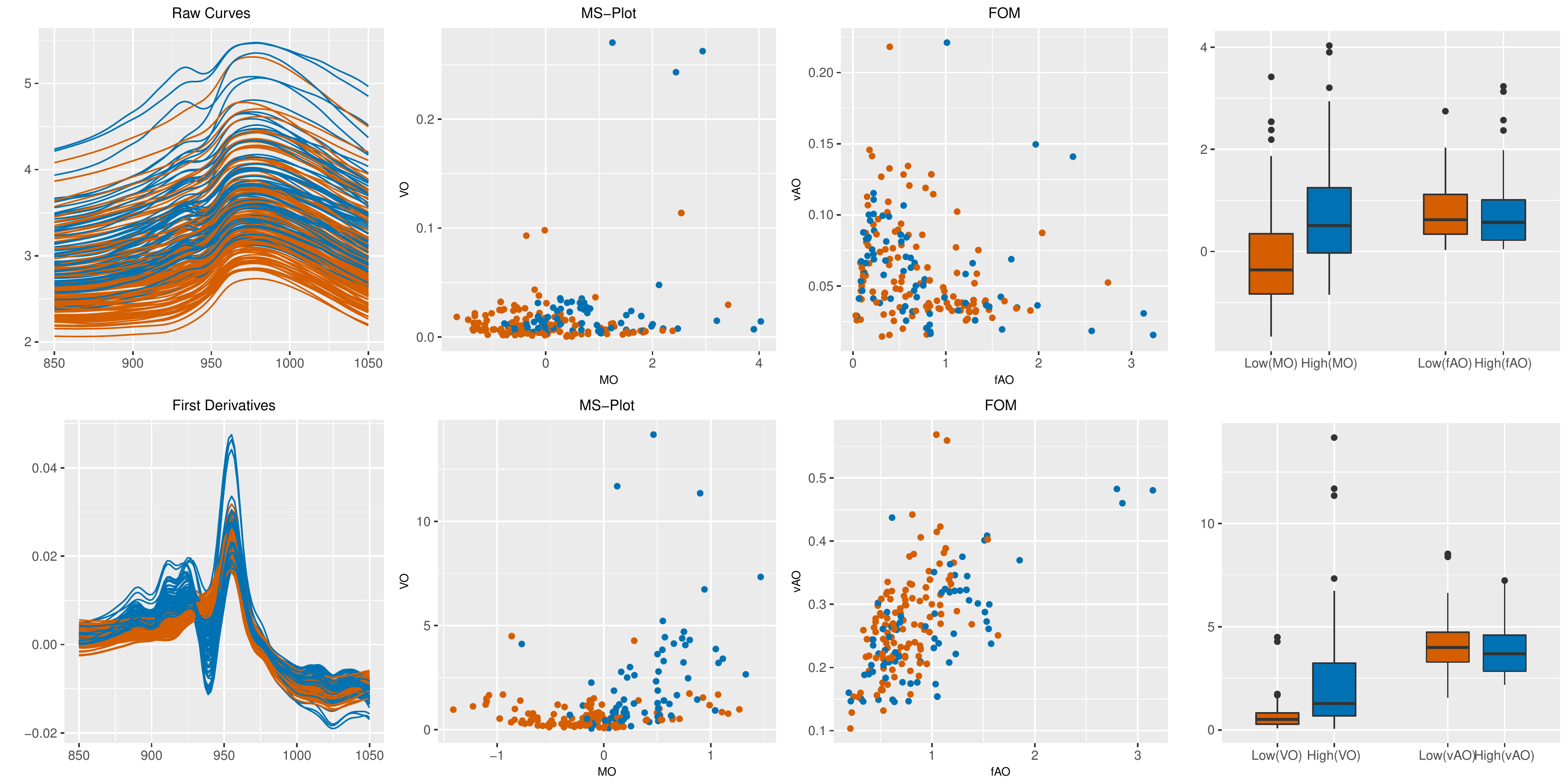}\\
\caption{Upper row: the spectrum curves and their MS-plot and FOM; lower row: the first-order derivatives of the spectrum curves and their MS-plot and FOM. The curves are divided into two groups according to fat-content percentage: greater than $20\%$ (blue) and less than $20\%$ (red). {The MS-plot better visualizes the differences of curves for both magnitude and shape.}}
\label{tecator-centrality}
\end{center}
\end{figure}

We illustrate the mean curves and the first-order derivatives of the two groups (first column, Figure \ref{tecator-centrality}), and apply the MS-plot and the FOM to assess their centrality (second and third columns, Figure \ref{tecator-centrality}).
For the mean curves, we observe that the blue curves are on average higher than the red curves.
Then, we check the corresponding MS-plot and FOM in the first row of Figure \ref{tecator-centrality} for this particular feature.
We find that the red and blue points tend to be located to the left and right of the MS-plot, respectively, whereas the two groups of points are well mixed in the FOM.
As an example, we provide the boxplots of MO and fAO for the two groups in the fourth column of Figure \ref{tecator-centrality}.
Apparently, the MO in the MS-plot describes the magnitude outlyingness of curves more accurately than the fAO in the FOM.
For the first-order derivatives in the second row of Figure \ref{tecator-centrality}, we reach a similar conclusion: the VO in the MS-plot describes the shape outlyingness of the curves more efficiently than the vAO in the FOM.
Overall, the above findings are consistent with our conclusions in Section~4.3.

\begin{figure}[b!]
\begin{center}
\includegraphics[width=16cm,height=8cm]{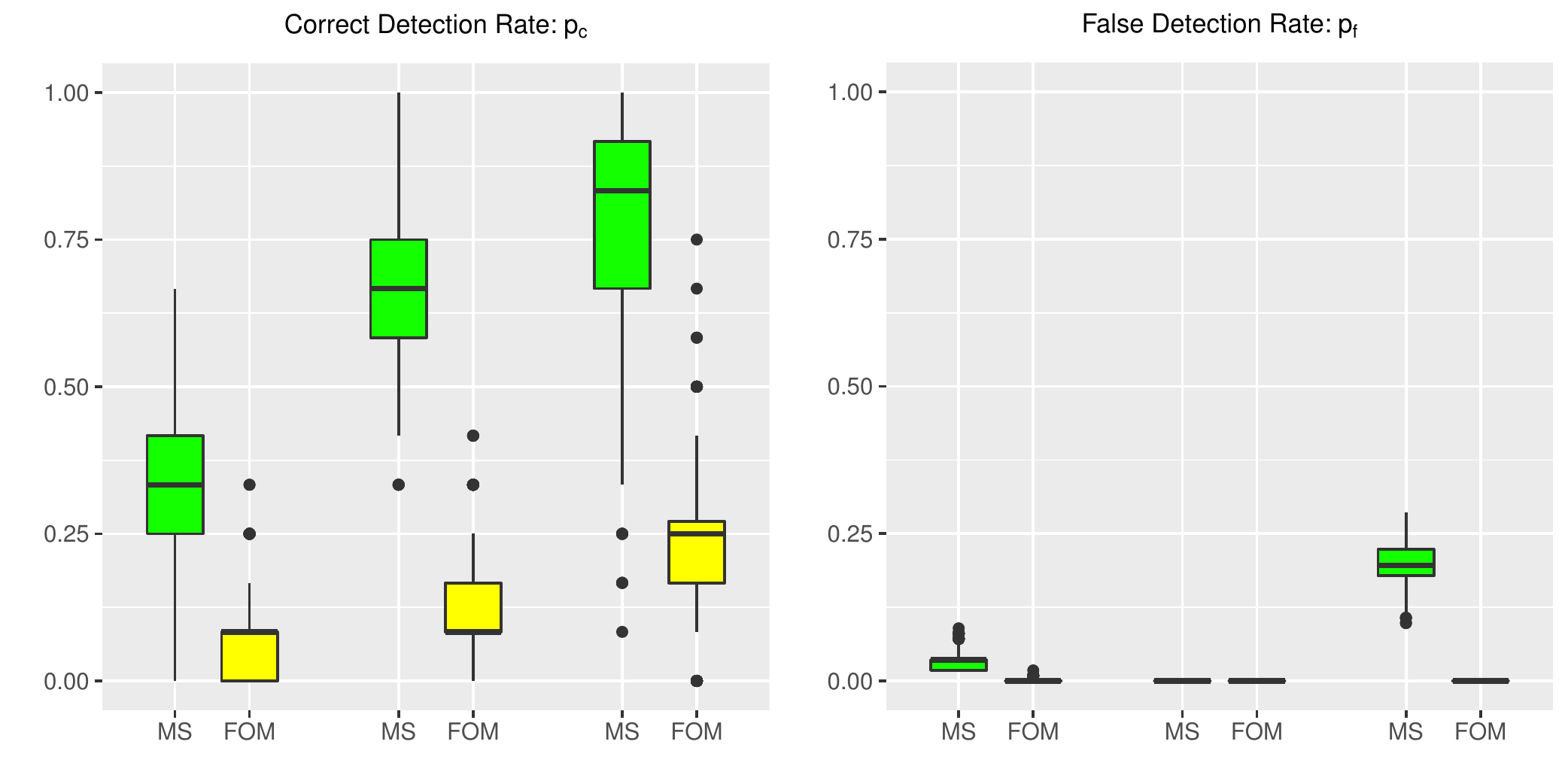}\\
\caption{Outlier detection results of the MS-plot (green) and FOM (yellow) for the Tecator data set. Left plot: $p_c$; right plot: $p_f$. First pair of box plots: results from the mean curves; second pairs: results from the first-order derivatives; third pair: results from the combination of mean curves and first-order derivatives. }
\label{tecator_detect}
\end{center}
\end{figure}

We also conduct outlier detection using the Tecator data.
Specifically, we treat the 112 samples with fat rates lower than $15\%$ as non-outlying curves, and randomly selected 12 curves with fat content higher than $20\%$ to treat as contaminated curves.
Thus, the contamination level is about $0.1$.
Then, we detect outliers using both the MS-plot and the FOM by combining the mean curves, the first-order derivatives, and the bivariate curves.
Following the simulation studies, we calculate $p_c$ and $p_f$ for each randomly chosen data set.
We repeat this procedure 1000 times and report the results in Figure \ref{tecator_detect}.
As illustrated, the $p_c$ of the MS-plot are always significantly higher than the $p_c$ of the FOM for all three types of data sources; the $p_f$ of the MS-plot are equal or a little larger than the $p_f$ of the FOM, due to the FOM's conservative outlier detection rule.
These findings are again consistent with our conclusions in Section 4.4; the MS-plot not only better describes the centrality of univariate functional data but also more efficiently detects various types of outliers.

\vskip 5pt
\subsection{Spanish weather data}

\begin{figure}[t!]
\begin{center}
\includegraphics[width=16cm,height=12cm]{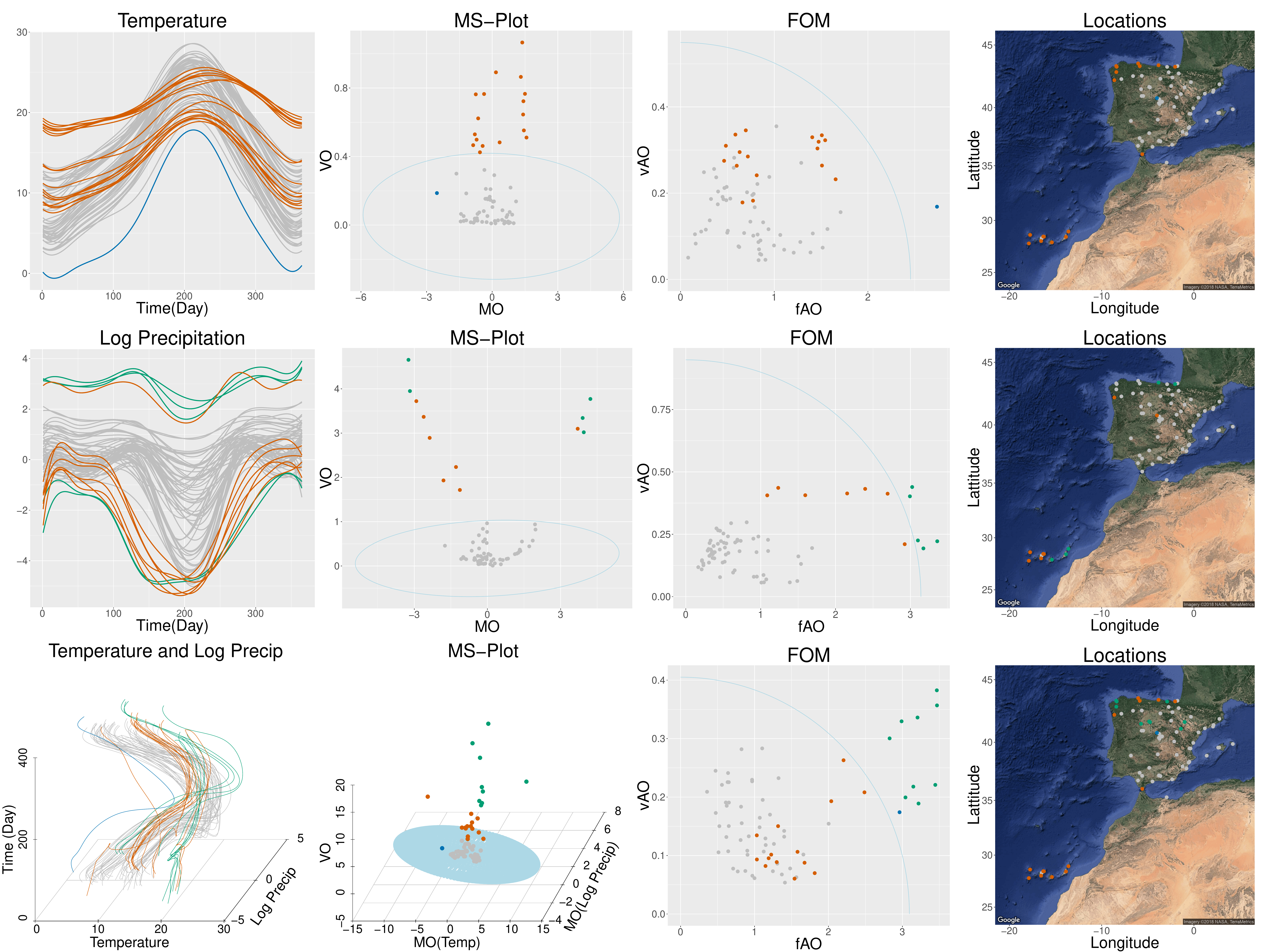}\\
\caption{First column: the smoothed temperature curves (top); the smoothed log precipitation curves (middle); the bivariate curves of temperature and log precipitation (bottom). Curves flagged as outliers by the MS-plot (red); curves flagged as outliers by the FOM (blue); curves flagged as outliers by both tools (green). Second and third columns: outlier detection results of the MS-plot and the FOM, respectively. Fourth column: the locations of the weather stations shown on a map of Spain.}
\label{weather-detect}
\end{center}
\end{figure}

The second data set includes averaged daily records from 73 weather stations in Spain during 1980--2009.
We obtain this data set from the R package \emph{fda.usc}.
At each station, daily temperature and daily log precipitation were recorded.
We smooth the data for analysis, and plot the curves both marginally and jointly in the first column of Figure~\ref{weather-detect}.
The MS-plot and the FOM are used to show centrality and to detect outliers from the curves in the second and third columns of Figure~\ref{weather-detect}.
The detection results are also illustrated, using the geographical information of each station, in the fourth column of Figure~\ref{weather-detect}. In addition, we provide two interactive MS-plot examples developed with the R package {\em plotly}; see the two links in Section 7. The first one is the MS-plot of smoothed temperature curves and the second one is the MS-plot of the bivariate curves of temperature and log precipitation.

No temperature curves are simultaneously detected as outliers by both methods: the MS-plot detects curves that can be regarded as shape outliers and the FOM identifies one curve that can be explained as a shifted outlier.
As shown in the fourth column of Figure~\ref{weather-detect}, the curves flagged as outliers by the MS-plot were all recorded at stations where the temperature changes more gradually compared with other stations,
either on the islands far away to the southwest of mainland Spain or on the Atlantic coast.
The curve flagged as an outlier by the FOM was observed at a station in the Madrid province at an altitude of 1894 meters; its average temperature is therefore the coldest, but its variation is similar to that of nearby stations.
These findings again confirm that the MS-plot is more sensitive to the shape outlyingness, which is harder to handle and of more interest in functional data analysis, rather than magnitude outlyingness.

For the log precipitation curves in the second row of Figure~\ref{weather-detect}, two groups can be clearly identified by either the curve patterns or the locations.
The first group contains four curves which are significantly higher than the others in the first plot; the second group includes the eight island weather stations, seen in the lower left corner of the fourth plot.
Both the MS-plot and the FOM clearly identify both groups, and the MS-plot further presents the different directions of deviation of the two groups, telling us whether the corresponding region is on average wetter or dryer than the rest of Spain.
Moreover, the MS-plot detected all the outlying elements in the two groups, but the FOM only detects some of them, which is a less desirable result.

For the bivariate curves in the third row of Figure~\ref{weather-detect}, the FOM detects the shifted outlier and identifies all of the first subgroup from the log precipitation data, but only part of the second subgroup;
the MS-plot recognizes the two subgroups completely as well as the stations of different temperature variations on the Atlantic coast.
Hence, the joint detection results are a combination of the results from the two marginal cases.

\vskip 12pt
\section{Discussion}
In this paper, we proposed the MS-plot for visualizing the centrality of functional data,
recognizing the influence of functional directional outlyingness on the outlyingness decomposition~\citep{dai2018directional}.
The MS-plot maps functional observations to multivariate points and can be applied to functional data with very general structures: both the response and support domains can be either one-dimensional or multi-dimensional.
Using both simulated data and practical examples,
the MS-plot was shown to be superior to the existing tools for visualizing both the level and the direction of magnitude outlyingness, for representing the various types of shape outlyingness, and for accurately detecting potential outliers.
For the convenience of implementation, we have developed a shinyapp; see the link in Section 7. Users can either assess the performance of the MS-plot for simulated data under various settings or analyze data sets uploaded from their local computers.

For spatio-temporal data, such as the Spanish weather data, we ignored the correlations between different locations and treated them independently.
Further improvements could be made by accounting for such correlations \citep{sun2012adjusted} when detecting outliers from the MS-plot, or even at the stage of defining functional directional outlyingness.
Also, we found that the distribution of $(\mathbf{MO}^{\rm T},{\rm VO})^{\rm T}$ is sometimes skewed in practical implementation because ${\rm VO}>0$, so a detection rule for skewed data \citep{sun2017robust} may further improve performance.

\vskip 12pt
\section{Supplemental Material}

\textbf{Plotly} Two interactive figures constructed with the R package {\rm plotly} are provided at {\textcolor[rgb]{0.00,0.00,1.00}{https://plot.ly/~langzi/117}} and {\textcolor[rgb]{0.00,0.00,1.00}{https://plot.ly/~langzi/119}}.

\noindent
\textbf{ShinyApp} The link for the shinyapp:  {\textcolor[rgb]{0.00,0.00,1.00}{https://langzizhiwen.shinyapps.io/msplot/}}

\noindent
\textbf{R-code} R-code \citep{Rcore} for constructing the MS-plot, the MS-plot array, and all the figures in this paper are provided. (codes.rar, GNU zipped tar file)

\noindent
\textbf{Additional Applications} Two applications on more generally structured functional data are demonstrated: the Cigarette data \citep{Chowdhury2016} are four-dimensional functional data recorded on a one-dimensional time interval; image data from a video \citep{rousseeuw2016measure} filmed by a static camera are three-dimensional functional data recorded on a two-dimensional space.

\begingroup
\setlength{\bibsep}{10pt}
\setstretch{0.5}
{
\bibliographystyle{asa}
\bibliography{visualization}

\begin{thebibliography}{46}
\newcommand{\enquote}[1]{``#1''}
\expandafter\ifx\csname natexlab\endcsname\relax\def\natexlab#1{#1}\fi

\bibitem[{Arribas-Gil and Romo(2014)}]{arribas2014shape}
Arribas-Gil, A. and Romo, J. (2014), \enquote{Shape outlier detection and
  visualization for functional data: the outliergram,} \textit{Biostatistics},
  15, 603--619.

\bibitem[{Chang et~al.(2017)Chang, Cheng, Allaire, Xie, and McPherson}]{shiny}
Chang, W., Cheng, J., Allaire, J., Xie, Y., and McPherson, J. (2017),
  \textit{shiny: Web Application Framework for R}, {R} package version 1.0.3.

\bibitem[{Chowdhury and Chaudhuri(2016)}]{Chowdhury2016}
Chowdhury, J. and Chaudhuri, P. (2016), \enquote{Nonparametric depth and
  quantile regression for functional data,} \textit{arXiv preprint
  arXiv:1607.03752}.

\bibitem[{Claeskens et~al.(2014)Claeskens, Hubert, Slaets, and
  Vakili}]{claeskens2014multivariate}
Claeskens, G., Hubert, M., Slaets, L., and Vakili, K. (2014),
  \enquote{Multivariate functional halfspace depth,} \textit{Journal of the
  American Statistical Association}, 109, 411--423.

\bibitem[{Cook and Swayne(2007)}]{cook2007interactive}
Cook, D. and Swayne, D.~F. (2007), \textit{Interactive and Dynamic Graphics for
  Data Analysis With R and GGobi}, New York: Springer.

\bibitem[{Croux and Ruiz-Gazen(2005)}]{croux2005high}
Croux, C. and Ruiz-Gazen, A. (2005), \enquote{High breakdown estimators for
  principal components: the projection-pursuit approach revisited,}
  \textit{Journal of Multivariate Analysis}, 95, 206--226.

\bibitem[{Dai and Genton(2018)}]{dai2018directional}
Dai, W. and Genton, M.~G. (2018), \enquote{Directional outlyingness for
  multivariate functional data,} \textit{Computational Statistics \& Data
  Analysis, {\rm to appear}}.

\bibitem[{Donoho(1982)}]{don0ho_1982breakdown}
Donoho, D.~L. (1982), \enquote{Breakdown Properties of Multivariate Location
  Estimators,} \textit{{\rm {P}h. {D}. qualifying paper, {H}arvard
  {U}niversity}}.

\bibitem[{Febrero et~al.(2008)Febrero, Galeano, and
  Gonz{\'a}lez-Manteiga}]{febrero2008outlier}
Febrero, M., Galeano, P., and Gonz{\'a}lez-Manteiga, W. (2008),
  \enquote{Outlier detection in functional data by depth measures, with
  application to identify abnormal NOx levels,} \textit{Environmetrics}, 19,
  331--345.

\bibitem[{Ferraty and Vieu(2006)}]{ferraty2006nonparametric}
Ferraty, F. and Vieu, P. (2006), \textit{Nonparametric Functional Data
  Analysis: Theory and Practice}, Springer.

\bibitem[{Filzmoser et~al.(2016)Filzmoser, Fritz, and Kalcher}]{pcaPP}
Filzmoser, P., Fritz, H., and Kalcher, K. (2016), \textit{pcaPP: Robust PCA by
  Projection Pursuit}, {R} package version 1.9-61.

\bibitem[{Genton et~al.(2014)Genton, Johnson, Potter, Stenchikov, and
  Sun}]{genton2014surface}
Genton, M.~G., Johnson, C., Potter, K., Stenchikov, G., and Sun, Y. (2014),
  \enquote{Surface boxplots,} \textit{Stat}, 3, 1--11.

\bibitem[{Gervini(2009)}]{gervini2009detecting}
Gervini, D. (2009), \enquote{Detecting and handling outlying trajectories in
  irregularly sampled functional datasets,} \textit{The Annals of Applied
  Statistics}, 3, 1758--1775.

\bibitem[{Gneiting et~al.(2010)Gneiting, Kleiber, and
  Schlather}]{gneiting2010matern}
Gneiting, T., Kleiber, W., and Schlather, M. (2010), \enquote{Mat{\'e}rn
  cross-covariance functions for multivariate random fields,} \textit{Journal
  of the American Statistical Association}, 105, 1167--1177.

\bibitem[{Hardin and Rocke(2005)}]{hardin2005distribution}
Hardin, J. and Rocke, D.~M. (2005), \enquote{The distribution of robust
  distances,} \textit{Journal of Computational and Graphical Statistics}, 14,
  928--946.

\bibitem[{Horv{\'a}th and Kokoszka(2012)}]{horvath2012inference}
Horv{\'a}th, L. and Kokoszka, P. (2012), \textit{Inference for Functional Data
  with Applications}, Springer.

\bibitem[{Hubert et~al.(2015)Hubert, Rousseeuw, and
  Segaert}]{hubert2015multivariate}
Hubert, M., Rousseeuw, P.~J., and Segaert, P. (2015), \enquote{Multivariate
  functional outlier detection,} \textit{Statistical Methods \& Applications},
  24, 177--202.

\bibitem[{Hyndman and Shang(2010)}]{hyndman2010rainbow}
Hyndman, R.~J. and Shang, H.~L. (2010), \enquote{Rainbow plots, bagplots, and
  boxplots for functional data,} \textit{Journal of Computational and Graphical
  Statistics}, 19, 29--45.

\bibitem[{Liu(1990)}]{liu1990notion}
Liu, R.~Y. (1990), \enquote{On a notion of data depth based on random
  simplices,} \textit{The Annals of Statistics}, 18, 405--414.

\bibitem[{L{\'o}pez-Pintado and Romo(2009)}]{lopez2009concept}
L{\'o}pez-Pintado, S. and Romo, J. (2009), \enquote{On the concept of depth for
  functional data,} \textit{Journal of the American Statistical Association},
  104, 718--734.

\bibitem[{L{\'o}pez-Pintado and Romo(2011)}]{lopez2011half}
--- (2011), \enquote{A half-region depth for functional data,}
  \textit{Computational Statistics \& Data Analysis}, 55, 1679--1695.

\bibitem[{Mahalanobis(1936)}]{mahalanobis1936generalized}
Mahalanobis, P.~C. (1936), \enquote{On the generalized distance in statistics,}
  \textit{Proceedings of the National Institute of Sciences of India}, 2,
  49--55.

\bibitem[{Mat{\'e}rn(1960)}]{matern1960spatial}
Mat{\'e}rn, B. (1960), \textit{Spatial Variation}, Springer.

\bibitem[{Pearson(1895)}]{pearson1895contributions}
Pearson, K. (1895), \enquote{Contributions to the mathematical theory of
  evolution. II. Skew variation in homogeneous material,} \textit{Philosophical
  Transactions of the Royal Society of London A: Mathematical, Physical and
  Engineering Sciences}, 343--414.

\bibitem[{{R Core Team}(2017)}]{Rcore}
{R Core Team} (2017), \textit{R: A Language and Environment for Statistical
  Computing}, R Foundation for Statistical Computing, Vienna, Austria.

\bibitem[{Ramsay and Silverman(2005)}]{ramsayfunctional}
Ramsay, J.~O. and Silverman, B.~W. (2005), \textit{Functional Data Analysis
  {\rm (second ed.)}}, Springer.

\bibitem[{Robbins(2012)}]{robbins2012creating}
Robbins, N.~B. (2012), \textit{Creating More Effective Graphs}, Wiley.

\bibitem[{Rousseeuw(1985)}]{rousseeuw1985multivariate}
Rousseeuw, P.~J. (1985), \enquote{Multivariate estimation with high breakdown
  point,} in \textit{{\rm In} Mathematical Statistics and Applications, Volume
  B {\rm (W. Grossmann, G. Pflug, I. Vincze and W. Wert, eds.)}}, Reidel,
  Dordrecht, pp. 283--297.

\bibitem[{Rousseeuw et~al.(2017)Rousseeuw, Raymaekers, and
  Hubert}]{rousseeuw2016measure}
Rousseeuw, P.~J., Raymaekers, J., and Hubert, M. (2017), \enquote{A measure of
  directional outlyingness with applications to image data and video,}
  \textit{Journal of Computational and Graphical Statistics{\rm, in press}}.

\bibitem[{Rousseeuw and Van~Driessen(1999)}]{rousseeuw1999fast}
Rousseeuw, P.~J. and Van~Driessen, K. (1999), \enquote{A fast algorithm for the
  minimum covariance determinant estimator,} \textit{Technometrics}, 41,
  212--223.

\bibitem[{Segaert et~al.(2017)Segaert, Hubert, Rousseeuw, and
  Raymaekers}]{mrfDepth}
Segaert, P., Hubert, M., Rousseeuw, P., and Raymaekers, J. (2017),
  \textit{mrfDepth: Depth Measures in Multivariate, Regression and Functional
  Settings}, {R} package version 1.0.4.

\bibitem[{Sievert et~al.(2017)Sievert, Parmer, Hocking, Chamberlain, Ram,
  Corvellec, and Despouy}]{plotly}
Sievert, C., Parmer, C., Hocking, T., Chamberlain, S., Ram, K., Corvellec, M.,
  and Despouy, P. (2017), \textit{plotly: Create Interactive Web Graphics via
  'plotly.js'}, {R} package version 4.7.0.

\bibitem[{Stahel(1981)}]{stahel1981breakdown}
Stahel, W.~A. (1981), \textit{Breakdown of covariance estimators}, Research
  Report 31, Fachgruppe f{\"u}r Statistik, ETH, Z{\"u}rich.

\bibitem[{Sun and Genton(2011)}]{sun2011functional}
Sun, Y. and Genton, M.~G. (2011), \enquote{Functional boxplots,}
  \textit{Journal of Computational and Graphical Statistics}, 20, 316--334.

\bibitem[{Sun and Genton(2012)}]{sun2012adjusted}
--- (2012), \enquote{Adjusted functional boxplots for spatio-temporal data
  visualization and outlier detection,} \textit{Environmetrics}, 23, 54--64.

\bibitem[{Sun et~al.(2017)Sun, Hering, and Browning}]{sun2017robust}
Sun, Y., Hering, A.~S., and Browning, J.~M. (2017), \enquote{Robust bivariate
  error detection in skewed data with application to historical radiosonde
  winds,} \textit{Environmetrics}, 28, e2431.

\bibitem[{Tarabelloni et~al.(2017)Tarabelloni, Arribas-Gil, Ieva, Paganoni, and
  Romo}]{roahd}
Tarabelloni, N., Arribas-Gil, A., Ieva, F., Paganoni, A.~M., and Romo, J.
  (2017), \textit{roahd: Robust Analysis of High Dimensional Data}, {R} package
  version 1.3.

\bibitem[{Tukey(1975)}]{tukey1975mathematics}
Tukey, J.~W. (1975), \enquote{Mathematics and the picturing of data,} in
  \textit{Proceedings of the International Congress of Mathematicians}, vol.~2,
  pp. 523--531.

\bibitem[{Unwin(2015)}]{unwin2015graphical}
Unwin, A. (2015), \textit{Graphical Data Analysis with R}, CRC Press.

\bibitem[{Vardi and Zhang(2000)}]{vardi2000multivariate}
Vardi, Y. and Zhang, C.-H. (2000), \enquote{The multivariate ${L}_1$-median and
  associated data depth,} \textit{Proceedings of the National Academy of
  Sciences}, 97, 1423--1426.

\bibitem[{Wickham(2010)}]{wickham2010layered}
Wickham, H. (2010), \enquote{A layered grammar of graphics,} \textit{Journal of
  Computational and Graphical Statistics}, 19, 3--28.

\bibitem[{Wickham(2016)}]{ggplot}
--- (2016), \textit{ggplot2: Elegant Graphics for Data Analysis},
  Springer-Verlag New York.

\bibitem[{Xie et~al.(2017)Xie, Kurtek, Bharath, and Sun}]{xie2017geometric}
Xie, W., Kurtek, S., Bharath, K., and Sun, Y. (2017), \enquote{A geometric
  approach to visualization of variability in functional data,} \textit{Journal
  of the American Statistical Association}, 112, 979--993.

\bibitem[{Yu et~al.(2012)Yu, Zou, and Wang}]{yu2012outlier}
Yu, G., Zou, C., and Wang, Z. (2012), \enquote{Outlier detection in functional
  observations with applications to profile monitoring,}
  \textit{Technometrics}, 54, 308--318.

\bibitem[{Zuo(2003)}]{zuo2003projection}
Zuo, Y. (2003), \enquote{Projection-based depth functions and associated
  medians,} \textit{The Annals of Statistics}, 31, 1460--1490.

\bibitem[{Zuo and Serfling(2000)}]{zuo2000general}
Zuo, Y. and Serfling, R. (2000), \enquote{General notions of statistical depth
  function,} \textit{The Annals of Statistics}, 28, 461--482.

\end{thebibliography}
}
\endgroup

\end{document}